# A Novel Hybrid Method for Network Anomaly Detection Based on Traffic Prediction and Change Point Detection

**Mouhammd Alkasassbeh**

*Department of Computer Science, Mutah University, Jordan, 61710, Jordan*
mouhammd.alkasassbeh@mutah.edu.jo

**Abstract:** In recent years, computer networks have become more and more advanced in terms of size, applications, complexity and level of heterogeneity. Moreover, availability and performance are important issues for end users. New types of cyber-attacks that can affect and damage network performance and availability are constantly emerging and some threats, such as Distributed Denial of Service (DDoS) attacks, can be very dangerous and cannot be easily prevented. In this study, we present a novel hybrid approach to detecting a DDoS attack by means of monitoring abnormal traffic in the network. This approach reads traffic data and from that it is possible to build a model, by means of which future data may be predicted and compared with observed data, in order to detect any abnormal traffic. This approach combines two methods: traffic prediction and changing detection. To the best of our knowledge, such a combination has never been used in this area before. The approach achieved a highly significant accuracy rate of 98.3% and sensitivity was 100%, which means that all potential attacks are detected and prevented from penetrating the network system.

## Introduction

With the increasingly rapid development of the digital world of information transmission and communication technology and with the adoption of such technology globally by individuals, institutions, governments and generally all sectors of society, the extensive availability of open source programs and management tools also fosters a constant threat of cyber-attacks, hacking and breaches of network security. Although security specialists proclaim that there is no "magic wand" that can be waved to instantly protect data on the Internet, effective actions can still be taken to prevent attacks on network security systems.

Traffic analysis has emerged as a promising approach to improving data protection. Traffic analysis is the process of monitoring, reviewing, modelling and analyzing traffic in a specific node or an entire network; to assess the network performance and security effectiveness. In this study, Simple Network Management Protocol (SNMP-MIB) is used as a data source in different scenarios that represent the most frequent and common types of attack encountered in real network operations. A model of network traffic using Artificial Neural Network (ANN) is developed in order to predict any new values entering the network imminently and comparing them with the real values.

Different techniques for network anomaly detection are mentioned in the literature; for example, signature-based detection, which works on patterns matching of the types of attacks that are already known. As this technique is effective by recognising the signatures of known attacks, new, unknown attacks are likely to be missed; therefore, regular updating is necessary. Other techniques, based on classifying abnormal network behaviour, need information about previous profile structures collected over a period of normal behaviour, to be effective. This technique is potentially highly effective in detecting new attacks without requiring specific, detailed knowledge. In our approach, we focus on the latter technique because it can detect both known and unknown attacks. It should be known that what is considered normal for one network could be abnormal for another and vice versa, therefore a dynamic and adaptive technique is required. In addition, our approach is simple to use and achieves a high degree of accuracy in detecting attacks. Speed is the most important factor in online detection, which is also a feature of our approach, as it combines traffic prediction using an ANN and change detection techniques using a control chart.

The rest of the paper is organized as follows: Section 2 discusses the work related to traffic prediction. The proposed approach is explained in section 3. The results and discussion are set out in section 4. Finally, our conclusions are presented in section 5.

## Related Work

Over the last decade, anomaly detection has appeared in the literature as an extensive area of study. Anomaly detection is a powerful data analysis technique, which is very useful for recognising network intrusions. Previously, most methodologies used in dealing with network anomalies consisted of general statistics comprising, for example, number of packets, packet length, IP and port numbers, etc. Another trend was also statistical, but used SNMP-MIB data as a data source. Some previous work on anomaly detection using SNMP-MIB will be briefly reviewed in this section.

The first attempt that used SNMP for network security is presented in (Cabrera *et al.*, 2002). The proposed methodology for the early detection of DDoS attacks was performed by applying statistical tests for causality in order to obtain MIB variables that present evidence of or signs to attacks.

In (Hoque and Chakraborty, 2015), authors used a statistical method reliant on the Kolmogorov Smirnov Test (KST) using SNMP-MIB data. Five main MIB groups were used (Interface, TCP, IP, ICMP and UDP). Their algorithm was tested against several types of DOS flooding attacks, for example: TCP SYN flooding, UDP flooding, ICMP PING flooding and IP flooding. Their algorithm achieved a lower false positive and negative rates in DOS attack detections.

Centralized intrusion detection systems are particularly vulnerable regarding scalability. To address this issue, in (AlKasassbeh, 2011), authors proposed a distribution system using Mobile Agent combined with a statistical method using SNMP-MIB data to tackle the scalability problem and the availability of network services.

In (Sangmee *et al.*, 2012), authors created a profile for network intrusion detection using MIB variables, with a specific focus on comparing current traffic data with traffic from the created profiles. This work achieved a high rate of detection accuracy by using a decision function to detect three different flood attacks: SYN, DNS and null scans.

As our approach is divided into two main parts, the second part being traffic prediction, it is worth mentioning some previous work in this area. Examples of used traffic prediction methods include Nonlinear Auto-Regressive Moving Average model (NARMA), Multi-Layer Perceptron (MLP), Support Vector Machine (SVM), Radial Basis Function Neural Network (RBF), Random Forest (RF), Random Tree (RT) …etc. Table 1 lists the most recent studies in the area of network traffic forecasting. What distinguishes these methods are the computational complexity and accuracy rate (the actual and predicted values should be very close). In other words, the accuracy, where the actual and predicted values, should be very close. In addition, the performance of algorithms plays an important role due to the shortage of resources.

## The Proposed Approach

This section demonstrates the proposed approach and the dataset being used. Figure 1 shows the overall architecture of the approach. In the first stage (building the learning model), the collected incoming traffic data is used as input to the Neural Network (NNARX) for traffic prediction. Section 3.1 provides information about the used data. In the second stage (Section 3.2), the NNARX model uses the normal network traffic to obtain the predicted traffic. The main concern is to score high accuracy (difference between the predicted and the actual traffic). Finally, the abrupt change detection step starts using a control chart model; this gives indications about any traffic outside the boundaries (abnormal traffic).

Table 1: A summary of recent studies in network traffic prediction methods

| Study | Method used |
|---|---|
| Nie *et al.* (2017) | Deep belief network and a Gaussian model |
| Wei (2017) | RBF neural network optimized by improved gravitation search algorithm |
| Sahrani *et al.* (2017) | Nonlinear Auto-Regressive Moving Average model (NARMA)+Multi-Layer Perceptron (MLP) |
| Akay and Akgöl (2016) | SVM, MLP, RBF, RF, RT, Reduced Error Pruning ( REPTree) |
| Katris and Sophia (2015) | Artificial Neural Networks, Multilayer Perceptron and Radial basis function+Autoregressive Fractionally Integrated Moving Average (FARIMA)+Holt–Winters, ARIMA/ Generalized Autoregressive Conditional Heteroscedasticity (GARCH)  and FARIMA/GARCH |
| Oliveira *et al.* (2014) | Multilayer Perceptron (MLP)+deep learning Stacked Autoencoder (SAE) |
| Chen *et al.* (2012) | Flexible Neural Tree (FNT)+Genetic Programming (GP)+Particle Swarm Optimization algorithm (PSO) |

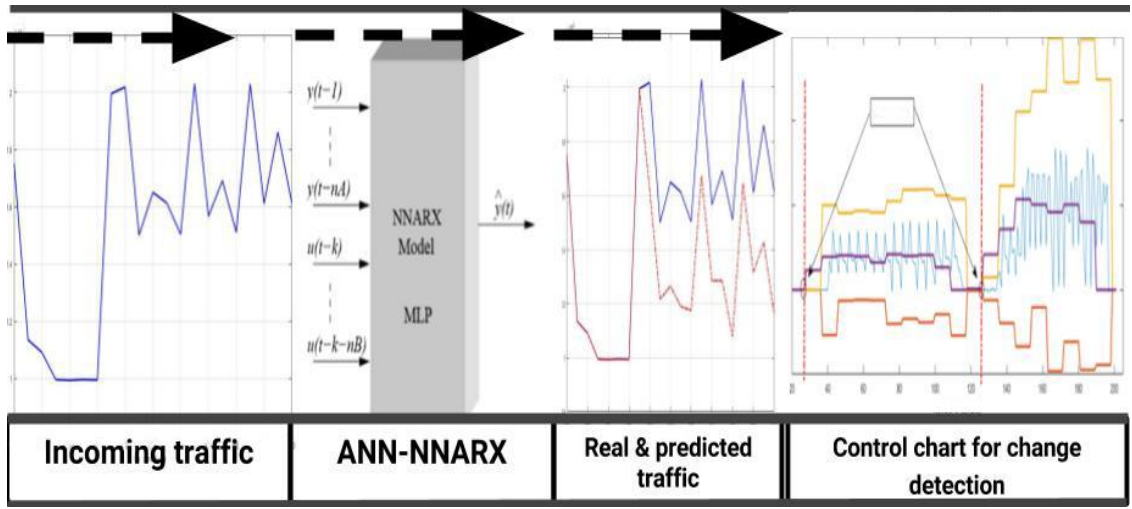

Fig. 1: General architecture of the proposed model

*The Dataset*

A network attack is any process used to perform malicious actions against any host inside a network that compromises the security of that network (Garber and Lee, 2000). In this study, the dataset in (Al-kasassbeh *et al.*, 2016) that was generated in our previous work was used to target DoS attacks. The objective of a DoS attack is to refuse requests from a legitimate user for services that the server is capable of providing. Such action can be achieved by flooding the chosen server with a high volume of traffic to consume all the server's resources, thereby preventing the server from responding to genuine requests. Such attacks can be generated either from a local or a remote network. A DoS attack is one of the most challenging threats that is difficult to assess and prevent (Ahmed *et al.*, 2016; Rao *et al.*, 2011). An even more serious threat than DoS is DDoS, which is a type of flooding attack from different nodes at the same time (Mirkovic and Peter, 2004). Six typical DoS flooding attacks described as follows.

TCP-SYN Attack: This attack affects the three-way handshake mechanism (SYN, SYN-ACK and ACK) operating between the host and the server at the establishment phase of TCP/IP protocol. During the process the attacker sends a SYN, the server on the other side responds to the SYN request by sending a SYN-ACK packet; meanwhile, the server stores and reserves all the resources and waits for an ACK from the client (the sender). While the server is waiting for the ACK, the request remains in the memory stack. The server will not receive ACK packets from the attacker and the attacker will send more SYN requests in a very short time so that the server's resources are drained to the extent that it cannot respond to any new requests (Alqahtani *et al.*, 2013).

UDP Flood Attack: This type of attack sends UDP packets to random ports of the victim server. When the server deals with these packets and discovers that the packets are empty, the server will then send back an error message through ICMP protocol to the sender. The server's resources, such as bandwidth, will be exhausted by the volume of useless packets and this hinders serving any other requests. A UDP flood attack is typically very effective in smaller networks (Salunke *et al.*, 2015).

*ICMP-ECHO Attack:* This type of attack floods the victim's bandwidth, preventing new connections from being initiated. The PING command is used to test whether or not the host is alive on the network so that when a device receives a PING request, it will automatically reply to the sender with a message informing of its status. This type of attack uses a trick of crafting a large number of ICMP packets with a spoof source IP address as the victim's IP address in order to reply directly to it later. These packets are then sent through a network broadcast address and are directed to send their replies to the same IP address at the same time. Eventually, the volume of reply messages will exhaust the victim's resources (Treseangrat, 2014).

*HTTP Flood Attack:* This attack targets a web server and consumes the victim's resources (e.g., memory, CPU, bandwidth). The attacker sends a great number of valid HTTP requests to GET or POST to the web server. Typically, these requests are generated by hosts called botnets. Each of the bots sends a large number of legal requests at the same time. In this case, if there is a great enough number of botnets, their request rates will be higher than the request rates of typical users. This attack may be one of the most dangerous threats, because it is hard to differentiate between normal and abnormal HTTP traffic (Zargar *et al.*, 2013).

*Slowloris Attack:* During such an attack, the attacker sends sessions with high workload requests by opening multiple connections to the victim's server and trying to keep these connections open as long as possible. In this case, the requests are partial HTTP requests. The attack lasts until all available sockets are taken and reserved by the HTTP requests and then the server freezes in response to any legitimate connections (Zargar *et al*., 2013).

*Slowpost Attack:* Similar to the previous attack but in which the attacker sends a complete, rather than a partial, HTTP header request, including the content length field in the post message body. The data fills the message body at the rate of one byte every two min and at the same time the server remains waiting for each message body to be completed, leading to denial of services (Zargar *et al*., 2013).

The source and type of data will now be clarified briefly. Management Information Base (MIB) was explored in earlier work (Al-Kasassbeh and Adda, 2009) in an investigation of the distributed model to exclude scalability problems in the network. Namvarasl and Ahmadzadeh (2014) proposed an intrusion detection system based on MIB and machine learning methods. Hoque *et al*. (2015) implemented a statistical method depending on the Kolmogorov Smirnov test for anomaly. Park and Kim (2008) proposed a lightweight algorithm for intrusion detection based on the correlation of MIB data. Other researchers used MIB variables for network problems with abnormality and network management. In this study, the dataset (Al-kasassbeh *et al*., 2016) has 34 MIB variables from 5 MIB groups in MIB-II as defined in (Rose, 1991): Interface, IP, TCP, UDP and ICMP group. In this research, we concentrated on the first group – Interface (IF). This clarifies information about all the interfaces of the node, for example; its interfaces number, physical address and IP address and other information related to this group.

*Network Traffic Prediction*

The main idea of the first part of this work is to train the ANN model using normal traffic and to achieve a very high accuracy rate of the prediction. If the difference between the incoming and predicted traffic is very close, then the network is operating normally. Otherwise, there is abnormal traffic coming into the node, in which case an alert should be issued to the manager or administrator. After building the ANN model and running it, it will always predict normal traffic; however, if there is an attack on the network, for example an ICMP_Echo attack, then the observed traffic will be different from the predicted traffic, with high convergence. The same will apply for other types of attacks. Where predicted traffic is close to actual traffic, in this case, the network is operating healthily and there are no attacks.

Network traffic prediction is essential for many applications that need to know the amount of traffic on its way to the network, such as anomaly detection, load balancer, network management and others. Using previous traffic to predict the future situation is one significant step towards dealing effectively with computer network issues. In addition, network traffic prediction is essential to reservation of resources in advance for the provision of a better service. Several techniques may be employed to achieve this task (Montgomery *et al*., 2015), such as Autoregressive (AR), Autoregressive Moving Average (ARMA), Autoregressive Integrated Moving Average (ARIMA) and Artificial Neural Network (ANN). Several studies discussed the use of either statistical models only or ANN; also the use of a hybrid model; that is something between a statistical model and ANN as in (C. a. Katris and Sophia, 2015). In other work that has used ANN, it was found that ANN was a competitive model to use and was superior to the classical regression model (Tang *et al*., 2016). In our work, we chose the ANN to build our model and use it for network traffic prediction as a first stage.

Neural Network is a very beneficial tool for solving prediction and modelling problems, because of its capability to learn and make generalizations. Learning relies on a set of training data provided from the problem's environment. The data can be used to tune the node weights by means of the learning algorithm. After the training process is completed, the network will be able to recognize (identify likenesses) or predict the output when new data are fed into it. The ANN can deal with non-linear, non-stationary data series as learning and input data. ANN is a network of interconnected nodes, derived from studies of biological nervous systems, in an attempt to create machines that operate in a similarly to the human brain (Rojas, 2013). ANN is mostly trained using learning algorithms for the purposes of learning and remembering a definite meaning or to classify patterns. Derived from biological neural networks, ANN is basically the simplest image of input and output units and algorithm for combining weights, as shown in Fig. 2.

where, $X_1$, $X_2$, ... $X_n$ represent the input data, $W_{nj}$ represents the weights, $Z_j = \sum x_i * W_{ij}$ and $f(Z_j)$ are the transfer function and the last output is $y_j$.

Multi-Layer Perceptron (MLP), also called Multi-layer Feed Forward (FF) network is the most common and widely used model. It comprises one input layer, one output layer and one or more hidden middle layers. Figure 3 shows a simple MLP NN with one hidden layer, in which a bias with initial weight value is added to the input layer. The input values are first fed into the input layer and are then multiplied by the weight values as they are passed from the input layer to the hidden layer. The network is fully connected as all input data are connected to all the nodes in the following layer. In each node of the hidden layer, multiplications are summed and then passed through the transfer function, such as nonlinear sigmoid function. The output layer also passes through linear sigmoid functions.

Different transfer functions are used in NNs. In this study, the most common transfer function was used: the logistic sigmoid function, which is defined in Equation 1 below:

$$f(x) = \frac{1}{1 + e^{\wedge}(-\beta x)} \tag{1}$$

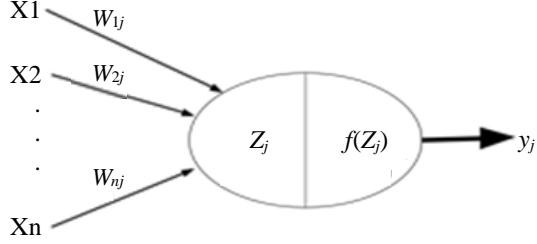

Fig. 2: General layout of artificial neural network

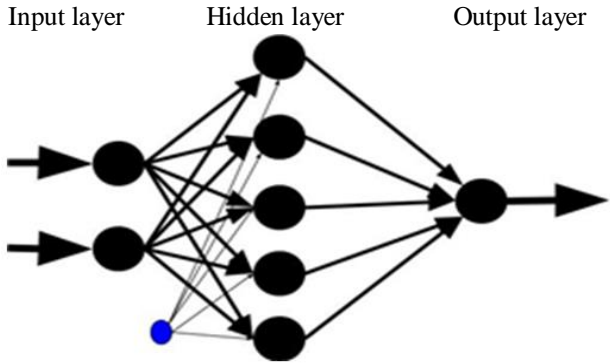

Fig. 3: Simple ANN architecture with three layers

Number of algorithms are used in MLP during the training phase. For example, Back Propagate algorithm (BP) is one of the most common neural network algorithms. The input data are entered continually to the ANN, where each output value is deducted to gain the required output and the error is also calculated. This error is then back propagated to the network to revise the weights so that better results with minimum error value are achieved. In each iteration, the desired output becomes closer to the actual output.

The use of an ANN to recognize dynamic nonlinear problems received more attention recently in cases where there is little previous knowledge of a problem. In addition, the system can be very complex, for example as in (Alkasassbeh *et al.*, 2013). Different types of ANN have different capabilities; a few types are capable of mapping nonlinear data to predict future ones using past input and/or output data. Examples of different ANN architectures may be found in (Commons *et al.*, 1991). In this study, our approach is based on the ANNARX Model which is presented in Fig. 4.

The proposed ANN architecture for predicting incoming traffic is presented in three layers, as shown in Fig. 5.

The input layer has eight input neurons, the hidden layer has seven neurons with nonlinear sigmoid function and the output layer has one linear output neuron to give the prediction of incoming traffic to assess the type of traffic based on predicted traffic.

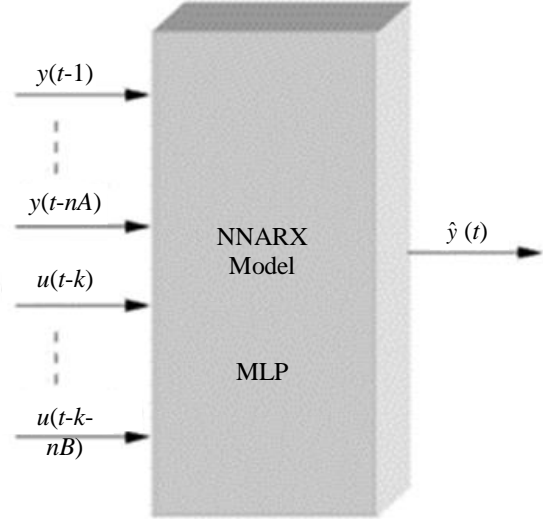

Fig. 4: The NNARX Model Structure (Norgaard et al., 2000)

In order to evaluate the model's performance in predicting incoming traffic in the normal case, the Mean Squares Error (MSE), Euclidean Distance (ED), Manhattan Distance (MD) and Mean Magnitude of Relative Error (MMRE) are measured. MSE, ED, MD and MMRE are calculated as shown in the following equations:

$$MSE = \frac{1}{n} \sum_{i=1}^{n} (y_i - \hat{y}_i)^2 \tag{2}$$

$$ED = \sqrt{\sum_{i=1}^{n} (y_i - \hat{y}_1)^2} \tag{3}$$

$$MD = \left( \sum_{i=1}^{n} \left| y_i - \hat{y}_l \right| \right) \tag{4}$$

$$MMRE = \frac{1}{n} \left( \sum_{i=1}^{n} \frac{\left| y_i - \hat{y}_i \right|}{I} \right) \tag{5}$$

where, $y$ and $\hat{y}$ are the observed traffic values and the predicted values based on the proposed model and $n$ is the number of instances used in the experiments, respectively.

*Change Detection Model*

A control chart is a basic method of statistical process control. For example, the progress of operations can be tracked by using statistics to ascertain whether there is an abnormal change in the process, thus determining a decision about taking necessary actions.

Normally, a control chart has a central horizontal line, denoting the mean of the sequence, with the Upper Control Limit (UCL) and the Lower Control Limit (LCL) defined by two other horizontal lines. For example, let $x_1$, $x_2$, $x_3$ … $x_n$ represent a sequence of data. In our case, it is the difference between predicted and observed traffic, therefore the general model of control charts in this case will be as the following; first, the mean of the sequence should be calculated as in Equation 6:

$$\bar{x} = \frac{x_1 + x_2 + x_3 + ... + x_n}{n} \qquad (6)$$

Calculation of the standard deviation can be made as follows:

$$\sigma = \sqrt{\frac{\sum_1^n (x_i - \bar{x})}{n-1}} \qquad (7)$$

The CL, UCL and LCL are calculated as follows:

$$UCL = \bar{x} + K\sigma \qquad (8)$$

$$CL = \bar{x} \qquad (9)$$

$$LCL = \bar{x} - K\sigma \qquad (10)$$

where, $k$ is the distance in two directions, upper and lower control limits, from the CL. In this research, the data are online, following on after each other and so piecewise windows were used, learning window and testing window. When data is incoming, it is necessary to wait for a period of time (window size). In our case, the window size used for learning and testing is 2.25 min (15 sec * 9 samples). After the windows filled with the samples, the ULC and LCL for the training window are calculated and these boundaries are applied to the testing window. As a result, any value that is outside these boundaries indicates that an error has occurred. As can be seen in Fig. 6, this method gives accurate results in recording the change in both the mean and the variance. The arrows and the vertical dashed red line in the Figure show the exact place of the error. The CL is also shown as being outside of the boundaries set by the (UCL and LCL).

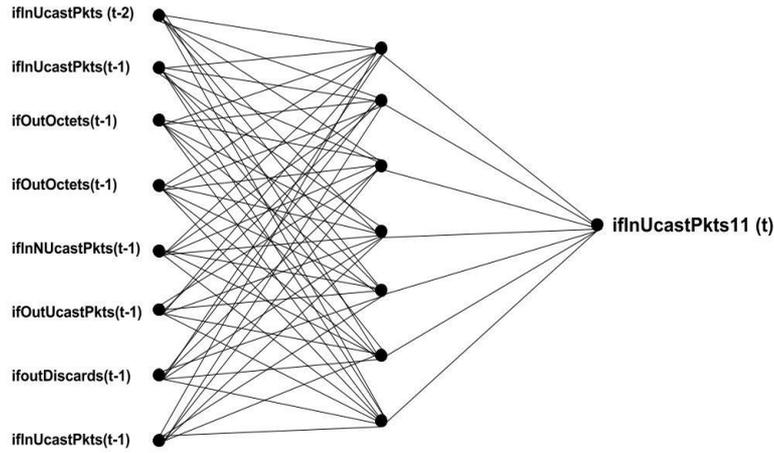

Fig. 5: The proposed ANNs architecture

$\times 10^4$                    Online control chart detection method

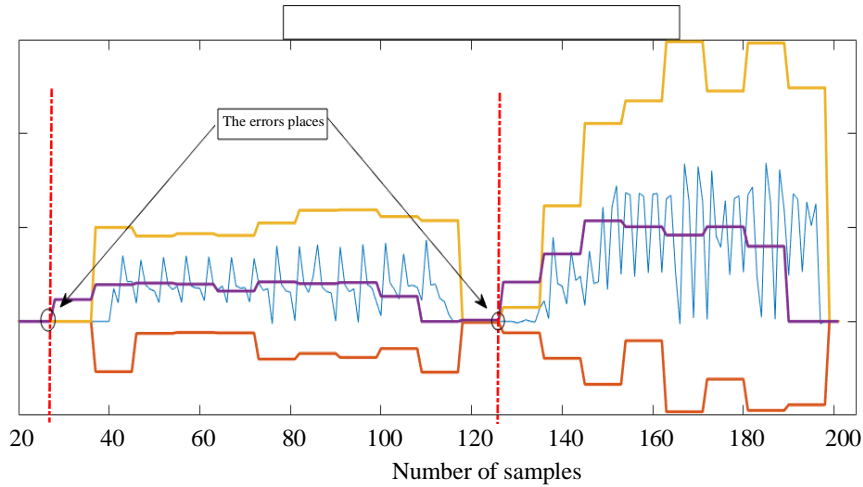

Fig. 6: Control chart example of two errors (TCP SYN and UDP flood)

## Experiments

In this section, the experimental results of the proposed approach will be described and explained in two main parts, the first discusses the experimental results for the ANN model and the second deals with the experimental results for the change detection model (control chart).

### Experimental Results for the ANN Model

The NNSYSID MATLAB Toolbox was used to develop the ANNARX, as provided by Magnus Nørgaard (2000). The ANNARX model is shown in Fig. 4 and the setup of the ANN parameters are listed in Table 4. The back propagation learning algorithm is used for the training stage of the ANNARX to reduce the convergence error.

The MIB variables that we used in our approach are used to train the ANNARX model. Data were collected every 15 sec in both the normal and abnormal mode. The 15 sec timing for polling data from the router was determined according to what was specified in many previous studies, such as (Thottan and Chuanyi, 2003). The dataset contained a total of 4998 MIB data records, as listed in Table 2.

Table 3 gives a basic summary of the statistics of the IF group, as used in this study. Some important information emerged from this table and highlighted the nature of the group used.

The proposed ANNARX model is run for a number of trials. It is found that the ANN model is able to achieve a high rate of accuracy with a minimal level of error. Figure 7 shows the performance of the model over the training normal traffic dataset. Figure 8 shows the results of testing the dataset. An evaluation of the model's performance, in terms of comparing predictions to observed data, is presented in Table 5.

It is clear that predicted traffic is close to the actual traffic. However, if an attack such as ICMP_Echo happens, the result of the model would be completely different, as shown in Fig. 9, where the predicted traffic is completely different from the observed traffic due to the attack behaviour. Table 6 presents the comparison results of the performance under normal and attack conditions. These show that the measurement metrics (MSE, MD, ED and MMRE) are very high. The same criteria are applied to the other attacks in the same manner but with different values, as each attack has its own style and behaviour. Figure 10 shows predicted and observed traffic for all attacks listed in Table 2.

### Experimental Results for the Change Detection Model

The control chart for change detection is an efficient method for identifying and recording all the attacks in our data; as there are seven attacks and the model captured all of them. The confusion matrix for our model is shown in Table 7.

Table 2: Traffic type and number of records generated

| Type of traffic | # of records |
|---|---|
| Normal | 600 |
| TCP-SYN | 960 |
| UDP flood | 773 |
| ICMP-ECHO | 632 |
| HTTP flood | 573 |
| Slowloris | 780 |
| Slowpost | 480 |
| Brute force | 200 |

Table 3: IF group summary and the standard deviation

| Attribute names | Min | Max | STD |
|---|---|---|---|
| ifInOctets | 1426588 | 4294415640 | 1233951794.00 |
| ifOutOctets | 161843 | 4294061114 | 1153395187.00 |
| ifoutDiscards | 0 | 196630 | 74228.9779 |
| ifInUcastPkts | 701369 | 243982848 | 58006834.50 |
| ifInNUcastPkts | 2735 | 35238 | 7936.895813 |
| ifInDiscards | 0 | 196630 | 74228.98972 |
| ifOutUcastPkts | 223076 | 93698308 | 23284377.61 |
| ifOutNUcastPkts | 796 | 8311 | 1784.778092 |
| tcpOutRsts | 1 | 4 | 1.086868566 |

Table 4: The adopted ANN model's parameters

| Parameters | Value |
|---|---|
| Number of layers | 3 |
| Number of neurons in the hidden layer | 7 |
| Number of neurons in the output layer | 1 |
| Number of iterations | 1000 |

Table 5: Evaluation criteria for normal traffic dataset

| Criteria | Training | Testing |
|---|---|---|
| MSE | 0.00790000 | 8.26270000 |
| MD | 0.06270000 | 1.42670000 |
| ED | 0.00000181 | 0.00001633 |
| MMRE | 0.00000000 | 0.00320000 |

Acknowledgement The authors feel greateful to the anonymous reviewer for their valuable comments and sugessions to improve the quality of paper and would like to thank them from core of the heart. Author's Contributions Sanjay Kumar: Conceptualization, Design and Analysis Drafting and Critical revision. Surjit Paul: Execution, Drafting and Revision. Dilip Kumar Shaw: Drafting the Manuscript and Revision. Ethics After publication of the paper, if we learn any sort of errors that changes the interpretation of the research findings, We are ethically obligated to promptly correct the errors in a correction, retraction, erratum or by other means.

| Criteria (normal) | Training Testing (attack) | |
|---|---|---|
| MSE | 0.00790000 | 72765000.00000 |
| MD | 0.06270000 | 2884.60000 |
| ED | 0.00000181 | 212910.00000 |
| MMRE | 0.00000000 | 0.02090 |

Table 7: Confusion matrix

| | Predicted attack | Predicted no attack |
|---|---|---|
| Actual attack | 7 | 0 |
| Actual no attack | 9 | 546 |

To measure the performance of our model, classification evaluation functions (Sensitivity and Specificity) are used. These are used mainly for binary classification and in our case it was either attack or no attack (abnormal traffic or normal traffic). Also, to evaluate the model based on the accuracy metric, equations for sensitivity, specificity and accuracy are as follows:

$$Sensitivity = \frac{TP}{TP + FN} \qquad (11)$$

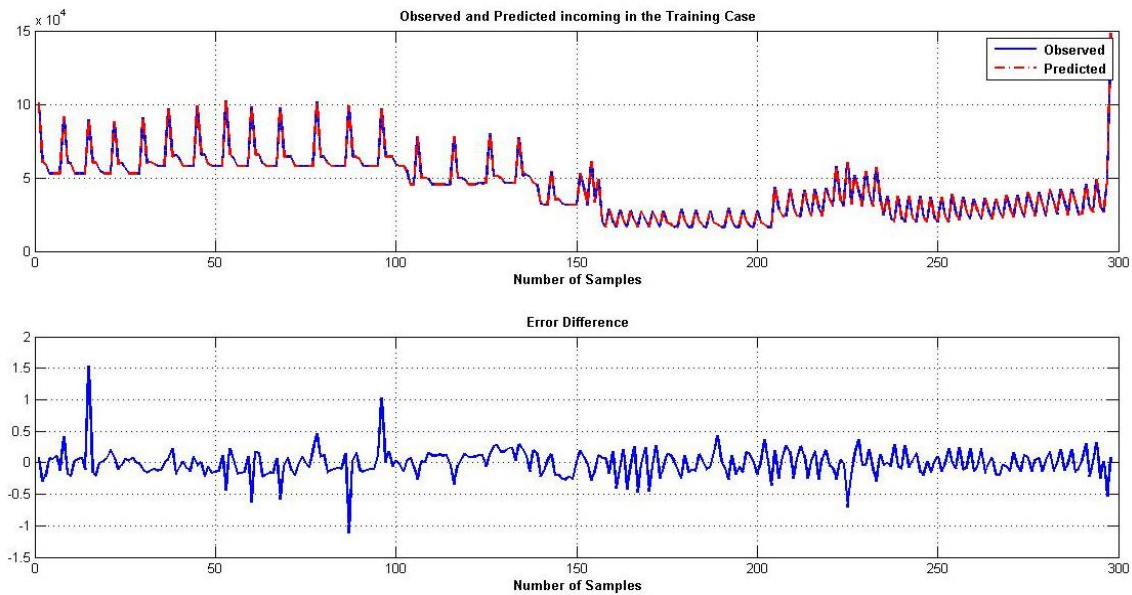

Fig. 7: Observed and predicted measurements for the incoming traffic ANN Model in the Training Case

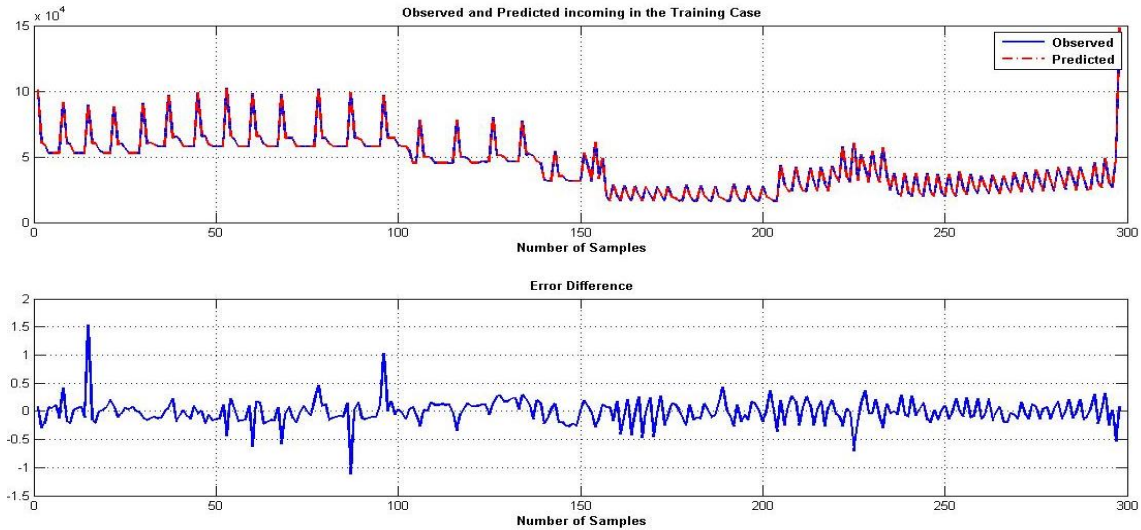

Fig. 8: Observed and predicted measurements for the incoming traffic ANN Model in the Testing Case

$$Specificity = \frac{TN}{TN + FP} \qquad (12)$$

$$accuracy(ACC) = \frac{TP + TN}{TP + TN + FP + FN} \qquad (13)$$

The True Positive (TP) samples are the correct cases classified positive and the False Positive (FP) samples are the negative cases that are classified positive. The True Negative (TN) samples are negative cases classified as negative. The False Negative (FN) samples are negative cases classified incorrectly as positive.

The model evaluation results show that the sensitivity, specificity and accuracy rates are 100, 98.3 and 98.3%, respectively. Consequently, the proposed model achieved excellent results.

## Conclusion and Future work

In this study, we demonstrated a novel and uncomplicated hybrid approach to capturing abnormal traffic at early stage with the purpose of helping network administrators to take preventive action quickly that will result in a more accessible and stable network. The accuracy of this approach is 98.3% and its sensitivity is 100%. It should be known that during the testing stage, our model captured all type of attacks (each with its own behaviour and style) with low false negative rates. Part of the future work, the false negative rate can be further reduced by improving and extending the MLP training stage on normal traffic. In addition, using dynamic window size of the control chart might give better results.

## Acknowledgement

The author feels extremely thankful to the anonymous reviewers, that work in this paper, for their respected comments and recommendations to increase the quality of this work.

## Ethics

There are no ethical issues with this paper.